# Journal of Geophysical Research: Space Physics

# Scale size-dependent characteristics of the nightside aurora

B. K. Humberset[1], J. W. Gjerloev[1,2], M. Samara[3], and R. G. Michell[3,4]

[1]Birkeland Centre for Space Science, Department of Physics and Technology, University of Bergen, Bergen, Norway, [2]The Johns Hopkins University Applied Physics Laboratory, Laurel, Maryland, USA, [3]NASA Goddard Space Flight Center, Greenbelt, Maryland, USA, [4]Department of Astronomy, University of Maryland, College Park, Maryland, USA

**Abstract** We have determined the spatiotemporal characteristics of the magnetosphere-ionosphere (M-I) coupling using auroral imaging. Observations at fixed positions for an extended period of time are provided by a ground-based all-sky imager measuring the 557.7 nm auroral emissions. We report on a single event of nightside aurora (∼22 magnetic local time) preceding a substorm onset. To determine the spatiotemporal characteristics, we perform an innovative analysis of an all-sky imager movie (19 min duration, images at 3.31 Hz) that combines a two-dimensional spatial fast Fourier transform with a temporal correlation. We find a scale size-dependent variability where the largest scale sizes are stable on timescales of minutes while the small scale sizes are more variable. When comparing two smaller time intervals of different types of auroral displays, we find a variation in their characteristics. The characteristics averaged over the event are in remarkable agreement with the spatiotemporal characteristics of the nightside field-aligned currents during moderately disturbed times. Thus, two different electrodynamical parameters of the M-I coupling show similar behavior. This gives independent support to the claim of a system behavior that uses repeatable solutions to transfer energy and momentum from the magnetosphere to the ionosphere.

## 1. Introduction

Only a few studies have attempted to address the spatiotemporal behavior of any electrodynamic parameter of the magnetosphere-ionosphere coupling, such as plasma convection and electrical conductances. The main reason for this shortcoming is the observational challenges. Observations must be made at a fixed point in the ionosphere at different times. This is not possible with a single satellite or rocket for which measurements are separated in both time and space. In stark contrast to the massive amount of single-satellite data, only a very limited number of multipoint satellite observations exist. For example, The Auroral Turbulence II sounding rocket mission [*Lynch et al.*, 1999], the Enstrophy sounding rocket mission [*Zheng et al.*, 2003], the CLUSTER II mission [*Forsyth et al.*, 2012], the Science and Technology 5 (ST 5) mission [*Gjerloev et al.*, 2011], and the Swarm mission [*Olsen et al.*, 2013]. The ST 5 mission is so far the most comprehensive multipoint data set of magnetic field perturbations collected by low Earth orbit satellites.

We can, however, utilize ground-based observations of auroral emissions since they also provide extended periods of continuous observations at fixed positions in a nonrotating reference system. Ground-based all-sky imagers are most often used for qualitative loose descriptions of the aurora (without any further data processing) in conjunction to other ground-based [e.g., *Dahlgren et al.*, 2012] or satellite observations. For example, the Time History of Events and Macroscale Interactions during Substorms (THEMIS) mission array of all-sky imagers are used to study the morphology of the aurora on a large scale (auroral oval) and in conjunction with the high-altitude THEMIS spacecrafts. Fixed point analyses by all-sky imagers have been used to understand the mechanism of fluctuating/pulsating aurora, such as tracing the characteristics of pulsating auroral patches in order to put constraints on the observed mechanism [e.g., *Humberset et al.*, 2016], or comparing the best correlation of the emission fluctuations to varying plasma properties at the magnetic equator [e.g., *Nishimura et al.*, 2010]. When it comes to spatiotemporal analyses, they are most often applied to narrow field-of-view imagers over small time intervals in order to investigate the small-scale flickering aurora [e.g., *Whiter et al.*, 2010], the very fast fluctuations superposed on the pulsating aurora [e.g., *Kataoka et al.*, 2012], or to reveal the nature and source of fine-scale structures in the aurora, such as by using different emission lines [e.g., *Dahlgren et al.*, 2016]. We will further focus on the lifetimes of the different mesoscale sizes of the nightside auroral display.

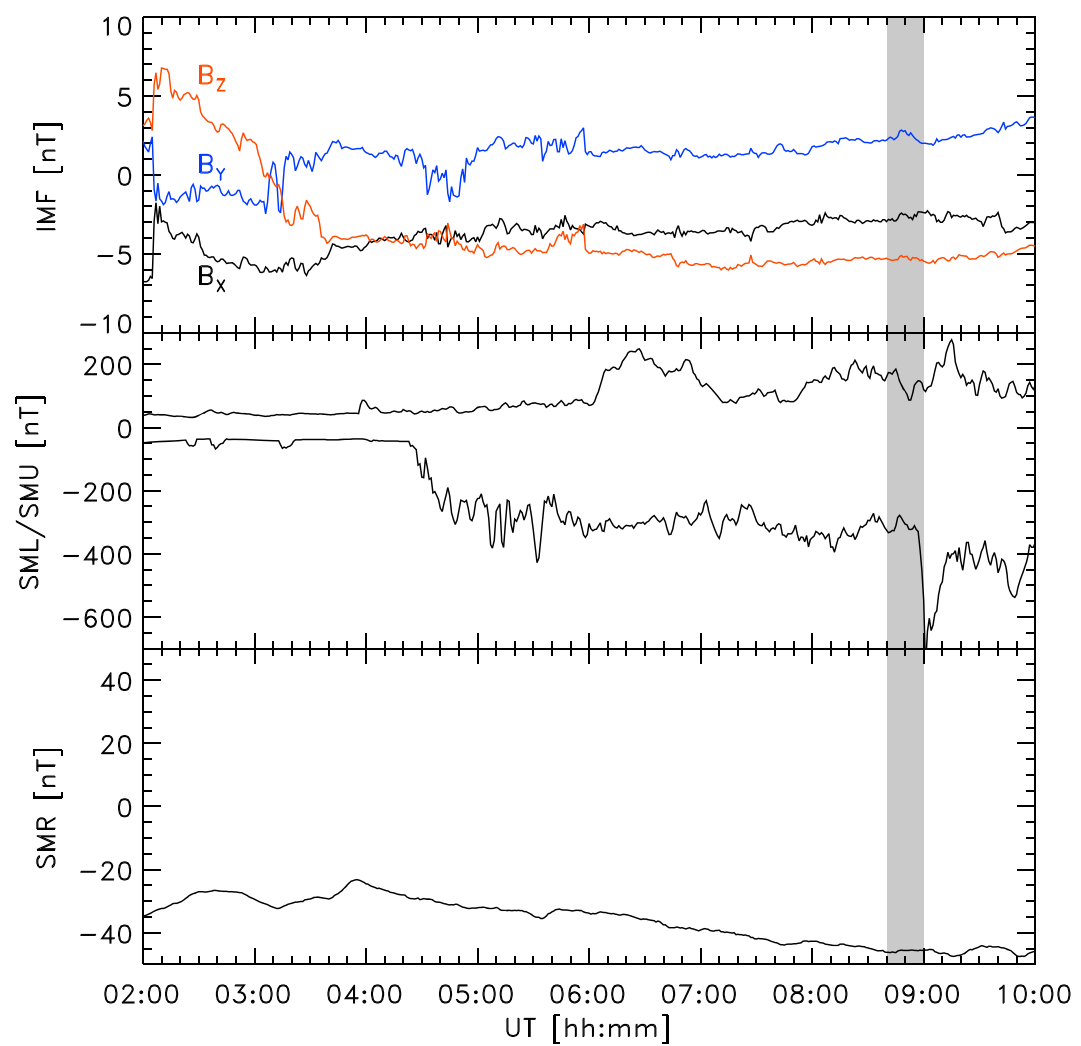

**Figure 1.** Geomagnetic indices (top) IMF, (middle) SML/SMU maximum westward/eastward auroral electrojet strength, and (bottom) SMR symmetric ring current index. The time of the event is highlighted in grey. We use the SuperMAG data set of indices and ACE IMF data which is propagated to the front of the magnetosphere (courtesy of Dr. James Weygand).

The purpose of this paper is to address the spatiotemporal characteristics of the magnetosphere-ionosphere coupling as observed by an all-sky imager measuring the 557.7 nm auroral emissions. In section 2 we describe the data and event; section 3 outlines the technique and methodology; in section 4 we show a typical example; section 5 shows statistical results; in section 6 we discuss the limitations and results; and finally in section 7 we summarize and draw conclusions.

## 2. Data and Event

The all-sky imager (ASI) utilized is located at Poker Flat at $-147.4°$ geographic longitude, 65.1° geographic latitude. The ASI uses a 557.7 nm narrow band filter and produces frames with 512 by 512 pixels at a frame rate of 3.31 Hz. Our event was recorded on 2 November 2011 at 08:41:06–08:59:58 UT (∼22:15 magnetic local time, ∼66° magnetic latitude). Throughout the event the sky is clear, the Moon is down and there are no artifacts such as street lamps.

Figure 1 shows the SuperMAG data set [*Gjerloev*, 2012; *Newell and Gjerloev*, 2011a, 2011b] of indices and propagated interplanetary magnetic field (IMF), which can be obtained through the SuperMAG website. The SuperMAG indices are derived according to the *AL*, *AU* and *SYM-H* indices, but utilize magnetometer observations from 100 or more sites instead of the 12 used in the official auroral electrojet indices and rather than six as is the case for *SYM-H*. Our event (highlighted in grey) occurred during an extended period of moderate activity driven by a southward IMF. The solar wind speed and the dynamic pressure (not shown) were fairly constant at ∼370 km/s and ∼1 nPa. The *SML* index of the maximum westward auroral electrojet strength shows 4 h of almost continuous activity starting with a substorm onset at 04:27 UT preceding our event, while the *SMR* indicates that there also is some ring current.

Our event is outlined with a keogram and five images in Figure 2. It starts with multiple slightly deformed east-west oriented arcs south in our field of view (FOV) until they fade about 08:46 UT. The entire FOV is then covered by relative dim and uniform arcs with slowly varying structures until a brightening and transition to highly structured rapidly changing arcs at about 8:49 UT. At around 8:55 UT highly deformed arcs extend along the magnetic latitudes in the center of the image while diffuse aurora can be found equatorward of the arc and a large part of the northern sky is dark. At about 8:57:30 UT we see that one of the arcs near the eastern horizon brightens and an auroral bulge spreads northwestward and saturates the ASI. This is likely the onset of an auroral substorm, which is also detected by the SuperMAG substorm database [*Newell and Gjerloev*, 2011a] at 08:57 UT.

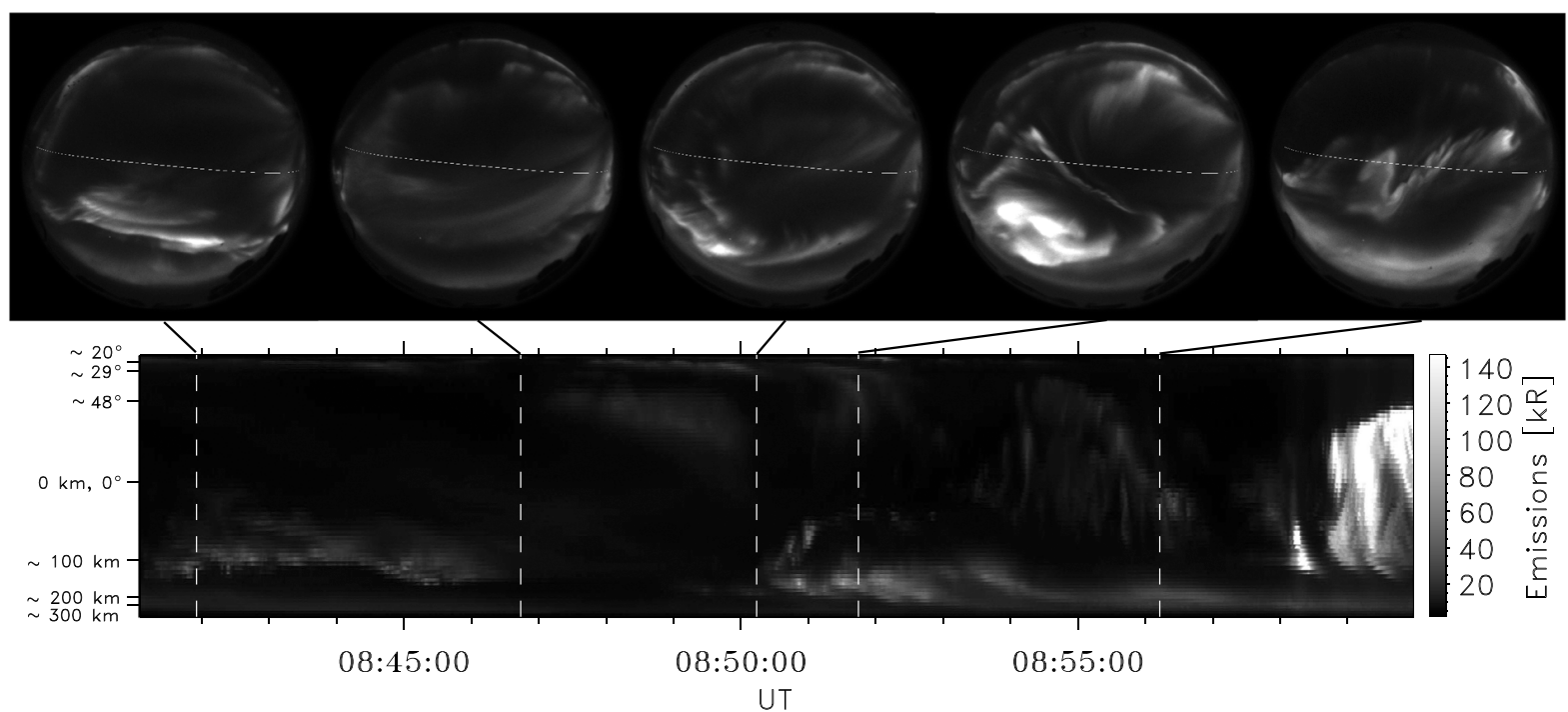

**Figure 2.** Keogram of the event and five example images. The y axis of the keogram shows the approximate distance from zenith and the elevation above the horizon. Note that the ASI images have a scaling of 0–70 kR, while the keogram has a scaling of 0–140 kR. This is done to avoid saturation of the keogram, while ensuring that the emissions are visible in the images. The magnetic east-west orientation is elucidated by the 66° magnetic latitude at 110 km altitude.

## 3. Technique and Methodology

Prior to the actual analysis we transform the ASI frames from the distorted fish-eye lens view to a Cartesian grid with uniform spatial resolution of 1.98 by 1.98 km, as described in *Humberset et al.* [2016]. The Cartesian grid is organized geographically as indicated on Figure 3. To avoid the most distorted limb pixels of the image, we find the characteristics within the center 507 by 507 km FOV (the square box). The spatiotemporal characteristics of the auroral emissions are found by performing a simple, yet robust, analysis that combines a two-dimensional (2-D) spatial Fast Fourier Transform (FFT) with a temporal correlation. This is a four-step process:

1. Perform a 2-D spatial FFT of each image and use a sweeping 2-D narrow Hanning-type band-pass filter to produce 256 filtered images.
2. Correct for Earth's rotation.
3. Calculate the correlation coefficient between two frequency-filtered images, $C = C(f, \Delta T)$, where f is the frequency of the band-pass filter, and $\Delta T$ is the time between two images.
4. Convert frequency to auroral scale size, $S$, to determine $C = C(S, \Delta T)$.

In step 2 we correct for the rotation of the all-sky imager with Earth. In the center of the image (65° geographic latitude) the Earth's rotation is 0.2 km/s eastward, assuming an emission altitude of 110 km [e.g., *Egeland and Burke*, 2013]. An auroral feature fixed in inertial space will therefore get an apparent westward velocity component of 0.2 km/s. To correct for this, we move the band-pass-filtered images 1 pixel (∼2 km) every ∼10 s eastward relative to the band-pass-filtered images at the start time ($T_0$) of the analysis. We only use the part of the image that is monitored throughout the time interval. For example, pixels that overlap for the entire event ($\Delta T = 18.9$ min) make up a FOV of 283 km in the east-west direction and 507 km in the north-south direction.

Step 3 calculates the linear Pearson correlation coefficient between two frequency-filtered images:

$$C(f, \Delta T) = \text{Correlate}(\text{Image}(f, T_0), \text{Image}(f, T_0 + \Delta T))$$

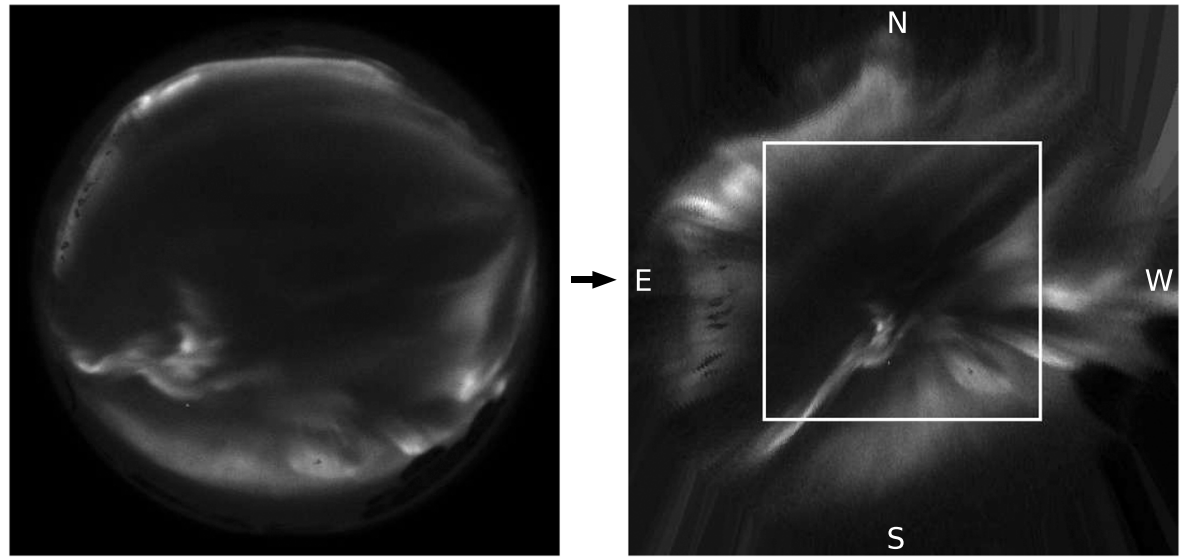

**Figure 3.** (left) The distorted ASI fish-eye view and (right) the resulting image when projected onto the Cartesian grid of 3.96 km² uniform pixel size. To avoid the most distorted limb pixels of the image, we find the characteristics within the center 507 by 507 km FOV (the square box). The Cartesian grid is organized with geographical north on top and west to the right.

Using all possible combinations of images and available frequencies would result in a total of $256 \cdot n \cdot \frac{(n-1)}{2} \simeq 1.8 \cdot 10^9$ correlation coefficients, where n = 3753 is the number of images in the event. To confine this overwhelming number, the statistical analysis is limited in two ways.

1. Because many of the high frequencies are equivalent to small and similar scale sizes, we ignore some of these higher frequencies. We use all lower frequencies (larger scale sizes). In short, we evaluate 35 of the 256 available frequencies inherent in the FFT.
2. We do not correlate all images with all other images. The first image is correlated to all subsequent images. Then we jump 10 images and correlate that to all subsequent images. Jump another 10 images and so on. This jump corresponds to ∼3 s.

These limitations reduce the number of calculations and affect the statistics that are used to derive the results without altering our findings. In parts of the event, especially in the last few minutes, the emissions are so bright that >50 pixels are saturated. These images are therefore not included in the analysis. With these limitations in mind we evaluate correlation coefficients for image separations of 1000 s or less, leaving us with a total of ∼$1.8 \cdot 10^7$ correlation coefficients. We linearly interpolate the resulting median correlation values to find the correlation coefficients for the frequencies that are not covered by the FFT.

In step 4 we determine the auroral scale size, S. The 2-D narrow band-pass filtering (step 1) is done in frequency space passing through frequencies that correspond to the scale size in question. This conversion between frequency and scale size is determined as

$$S = s(f_x) \cdot s(f_y) = \frac{512}{2f_x} \cdot \frac{512}{2f_y} \cdot 3.92 \ \text{km}^2 \qquad (1)$$

The scale size (S) is an area, where 512 is the number of pixels in the image in the x and y directions and 3.92 km² is the pixel size, again assuming an emission altitude of 110 km. The frequencies $f_x$ and $f_y$ in the x and y direction, respectively, are inherent in the FFT. In this way we can determine the scale size-dependent variability of the auroral images, $C = C(S, \Delta T)$. The scale size-dependent amplitude $A = A(S)$ is determined from the complex spectrum as the square root of the power.

## 4. Typical Example

This section is intended to illustrate the technique outlined in the previous section as well as to show a typical example supporting the statistical results presented in the next section.

Figure 4 shows examples of four image separations $\Delta T$; 0.3 s, 2 s, 10 s, and 60 s. The panels show the respective images as well as their absolute difference and scale size-dependent correlations, $C(S,0.3)$, $C(S,2)$, $C(S,10)$, and $C(S,60)$ calculated from the corresponding sets of band-pass-filtered images. The images are scaled according to a constant color scale, while the difference for visual purposes is scaled according to the maximum difference, which in general are increasing with the image separation. Again, recall that the Earth's rotation has been removed and thus the analysis is performed in an inertial reference frame. The absolute difference is purely for visual inspection, not for actual analysis, but it clearly shows the largest spatial scale sizes that have changed for the respective image separations. Figure 4 (top) shows the smallest possible image separation (0.3 s) in the data set. Even after close inspection, the images look the same. However, their difference shows salt and pepper, meaning small auroral filaments have changed. It can be argued that this is only noise, but the higher intensity in the location of the bright aurora suggests that the structures are real. Moving toward larger image separations, it is easier to see that the images are different and the difference comprises increasingly larger auroral features. The correlation analysis supports this finding; As Δ*T* increases, the correlation decreases for increasingly larger scale sizes per square kilometer. When the image separation has reached a minute, we clearly see that the large auroral form has started to change and the correlation analysis shows that auroral forms of almost all scale sizes have changed. It is clear that the small scale sizes are more variable.

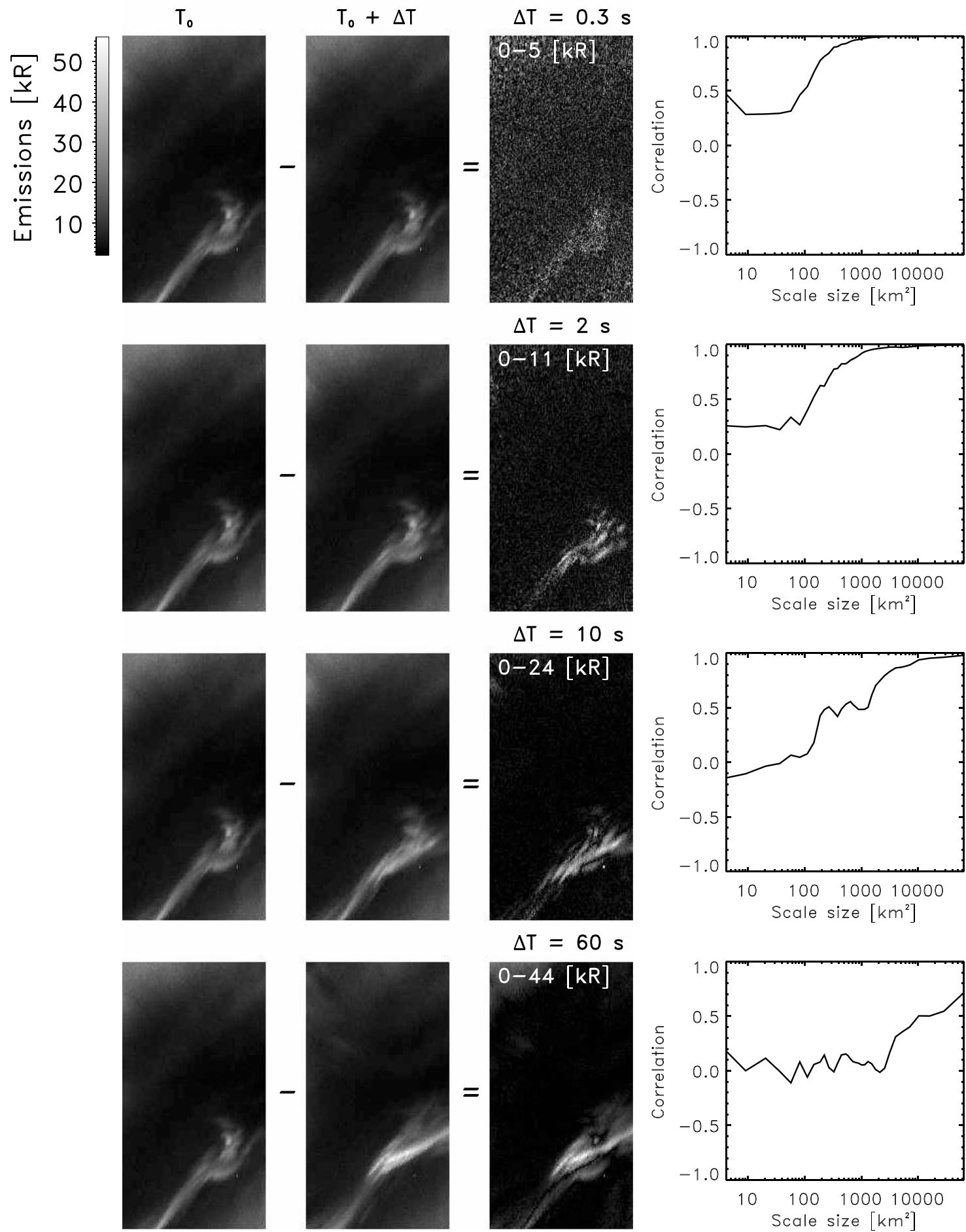

**Figure 4.** Examples of four-image separations ΔT; 0.3 s, 2 s, 10 s, and 60 s. The panels show the respective images as well as their absolute difference and scale size-dependent correlation. As ΔT increases, note how the correlation drops off for larger scale size as larger auroral forms show up in the difference between the images that are compared.

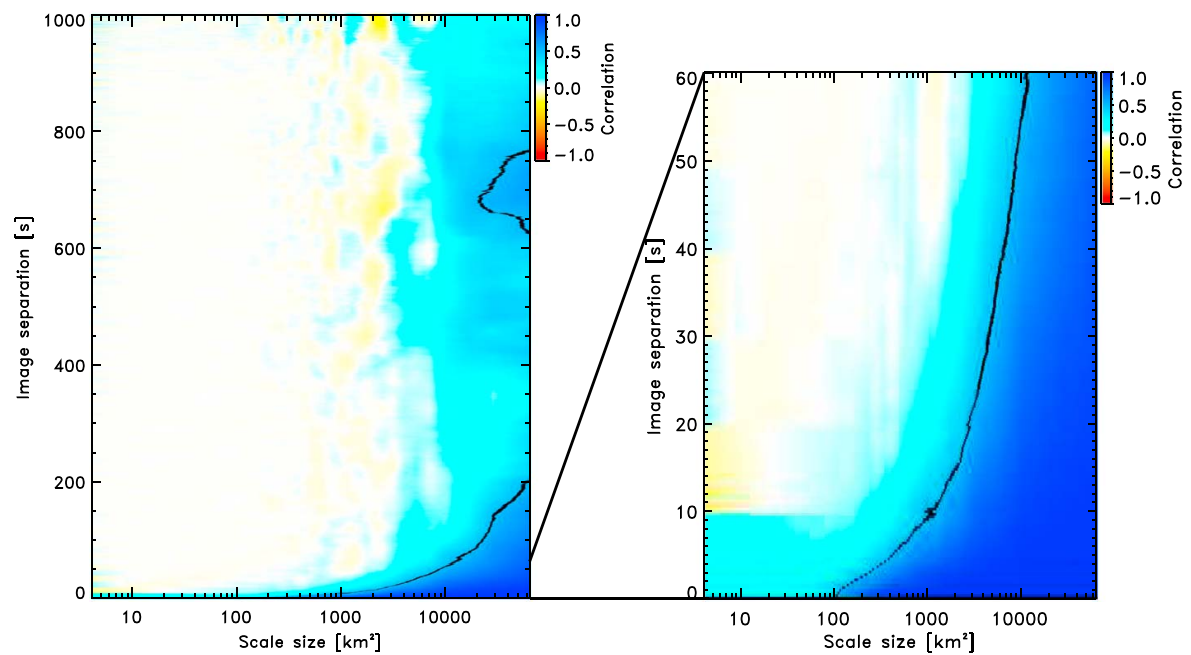

**Figure 5.** Correlation as a function of scale size (*S*) and image separation (Δ*T*). The artifact clearly visible for the small scale sizes is due to the stepwise correction of Earth's rotation for every 10 s image separation. Note the well-organized regions of high and low correlation.

## 5. Results

The 2-D correlation distribution $C = C(S, \Delta T)$ is shown in Figure 5. The result is a fairly organized pattern of high correlation and low correlation. For small scale sizes and large image separation the two sets of observations are uncorrelated while short image separation and large scale sizes are highly correlated. We use a correlation of 0.5 to provide a loose separation boundary between regions of "high correlation" and "low correlation" (the black lines in Figure 5). For example, we could assume that correlation values >0.5 means that the auroral forms have not changed much and can be considered relatively stable. This would then imply that auroral forms of 1000 km$^2$ can be considered as static on timescales of about 10 s or less, while a 10,000 km$^2$ auroral form can be considered as static on timescales less than a minute. Moving the threshold for what is considered relatively static aurora to correlation values >0.6 or another reasonable number would shift the timescales to lower values, but the overall finding is unaltered. The horizontal lines made by shifts in the correlation values are likely a result of the technique used to correct for Earth's rotation, which is not continuous but moves the images relative to each other 1 pixel every 10 s. Figure 5 also shows that the largest scale sizes are relatively well correlated for an interval of longer image separations. The magnetosphere-ionosphere system (as observed by 557.7 nm emissions) appears to display a scale size-dependent variability where the small scale sizes are more variable and the larger scale sizes more stable.

## 6. Discussion

We interpret the intriguing coherence pattern shown in Figure 5 as indicative of an underlying magnetosphere-ionosphere (M-I) system behavior. It may sound trivial that small scale sizes change more rapidly than larger scale sizes, but that implies a systematic behavior of the M-I system. However, before we interpret our results we address the inherent limitations of the data, methodology, and technique.

### 6.1. Inherent Limitations

The three main assumptions are as follows: (1) The spatiotemporal characteristics of the emissions do not change during the 19 min interval; (2) artifacts due to the 2-D spatial FFT filter are negligible; and (3) conclusions are limited to imager capabilities.

When calculating the correlation coefficient, $C = C(S, \Delta T)$, we assume that this parameter is only a function of $S$ and $\Delta T$. In other words, $C(S, \Delta T, T1) = C(S, \Delta T, T2)$, or the characteristics, are time independent. This is also the case for the amplitude: $A(S, T1) = A(S, T2)$. The validity of this assumption will be investigated in sections 6.3 and 6.4.

The use of a 2-D spatial FFT band pass to analyze scale sizes of course has its limitations and effects that we should be aware of. Equation (1) in step 4 of the technique explains that for each band-pass-filtered image, the band-pass filter all frequency combinations $f = f_x \cdot f_y$ that correspond to the area $S = s(f_x) \cdot s(f_y)$. This is not unambiguous compared to working with one-dimensional band-pass filtering, because we might filter very different shapes. The visual aurora appears in a large variety of forms, such as patches, elongated arcs, and arcs that are deformed by spirals and folds. One scale size and band-pass-filtered image can therefore contain both quadratic shapes ($s(f_x) = s(f_y)$) and very elongated shapes ($s(f_x) \neq s(f_y)$) of the same area, if they are present in the original image. Also the regions of dark skies between the aurora are embedded in the scale sizes. The use of a narrow band pass can of course produce artifacts and uncertainties to the scale sizes. However, these are likely minimized because we in the end are interested in the change between two images filtered with the same band pass. This change is quantified using the linear Pearson correlation coefficient, which has the embedded mathematical property that it is invariant to a dimming or brightening but highly sensitive to distortions in the image. The scale sizes are validated by the typical example (section 4), which shows that the correlation falls off for the same scale sizes that are visible in difference between the images. For instance, in the typical example of a 60 s image separation we clearly see that the correlation of the largest scale sizes falls off and thus responds to the overall change in the auroral arc and not as much by relatively stable large region of dark sky poleward of the arc. Further, the typical example of 60 s image separation shows a scale size-dependent correlation coefficient in excellent agreement with Figure 5. Thus, the artifacts due to the use of a narrow 2-D spatial FFT filter to evaluate the change in auroral images are negligible.

An all-sky imager provides good coverage of the night sky and allows monitoring of rapidly evolving features, but the FOV limits the observations to mesoscale features and the spatial resolution cannot resolve fine-scale features. A more concerning problem, however, is the inherent distortions of an ASI. An ASI is a very wide FOV imager that obtains column-integrated measurements of the auroral emissions. The distortions are due to a variety of different issues which have been discussed in past technical papers. The main issue is that the emissions come from a range of altitudes and are primarily due to precipitating electrons which effectively paint the field lines as they collide with the neutral atmosphere. The ASI obtains column-integrated emissions which typically originate from a range of altitudes and magnetic field lines. Additionally, there are distortions due to optics and elevation (look direction).

The question is whether smearing affects our ability to answer the stated science objective. While there is no robust way to remove smearing, we mediate it by only using the center 507 by 507 km FOV (see Figure 3). The resampled image has a pixel size of 1.98 km and based on the size of the largest pixels in the fish-eye view that are projected onto the resampled grid, we assume that the smearing is about 5 km. The spatiotemporal characteristics of features with scale sizes smaller than 5 km can therefore not be resolved.

### 6.2. What Is the Source of the Variability?

A velocity component of the auroral arcs will cause a phase shift of the band-pass-filtered images, which will affect the correlation analysis. We have effectively found the scale size-dependent total derivative:

$$\frac{dI}{dt} = \frac{\delta I}{\delta t} + \mathbf{U} \cdot \nabla I$$

where $I$ is the auroral emissions and $\mathbf{U}$ is the velocity of the auroral arcs and forms. This means that the observed variability in the auroral display can be due to changes in the partial derivative with respect to time as well as any movement of the auroral arcs. Separating these terms in order to determine the cause of the observed dI/dt is not straightforward.

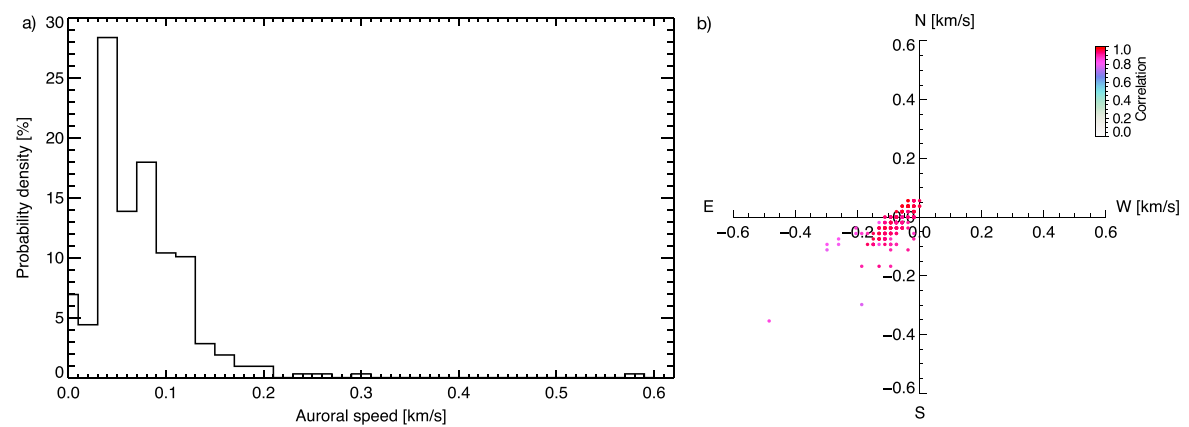

**Figure 6.** (a) The auroral speed probability from a simple cross correlation between images separated by $\Delta T = 10$ s. (b) The velocity vectors on a geographic grid with the correlation value denoted by the color scale.

To determine the relative importance of these two source terms (partial derivative and convective term), we performed a simple cross correlation between images separated by $\Delta T = 10$ s. The 2-D spatial shift can be related to a spatial vector, and since we know the time elapsed we straightforwardly calculate the velocity vector (**U**). The weakness of this technique is, of course, that the cross correlation is largely driven by the large scales and thus it is assumed that all scale sizes are moving with the same velocity. This is not necessarily true. Figure 6 shows the resulting velocity vector (**U**) and the speed probability. Using simple estimates, we get the convective term to be on the order of 0.3 kR/s and the total derivative to be on the order of 2.4 kR/s. Thus, the partial derivative must be on the order of 2.1 kR/s leaving us to conclude that we can largely ignore the convective term.

### 6.3. Controlling Parameters

The single event studied in this paper provides an opportunity to prove the technique and provides a glimpse of the M-I system characteristics, but it does not allow general conclusions. Statistical studies are needed to address how the characteristics of the auroral emissions may be dependent on local time, geomagnetic conditions, seasonal effects, or any other controlling parameters. During our event the solar wind driver is fairly constant but near the end of the interval a substorm onset takes place. This saturates the imager which forces us to terminate the analysis. Thus, we can assume fairly constant geomagnetic conditions, solar wind driver conditions, and since the event is fairly short, the geomagnetic location can be assumed constant.

Figure 5 was derived from the entire interval and thus represents an average result. We may question the assumption that the spatiotemporal characteristics are fairly constant. If we separate the movie into smaller intervals we can test this assumption. We identify two intervals for which we loosely define the aurora as visually (a) "less variable" and (b) "more variable." In interval a (08:46:25 – 08:48:55 UT) the entire FOV is covered by relative dim and uniform arcs with slowly varying structures. Interval a has a median minimum intensity of ∼3 kR and maximum of ∼35 kR. In interval b (08:55:04 – 08:57:35 UT) highly deformed arcs extend along the magnetic latitudes in the center of the image while diffuse aurora can be found equatorward of the arc and a large part of the northern sky is dark. Interval b has a median minimum intensity of ∼2 kR and maximum of ∼52 kR.

Figure 7 shows the correlation as a function of scale size and image separation for interval a of less variable aurora (Figure 7a) and interval b of the more variable aurora (Figure 7b). The characteristics from interval a has higher correlations (more blue) for scale sizes of ∼2000 km$^2$ to ∼5000 km$^2$, and perhaps more surprising, lower correlation for scale sizes >9000 km$^2$ and <800 km$^2$ compared to interval b. For example, we could assume that correlation values >0.5 means that the auroral forms have not changed much and are relatively stable. This would then imply that auroral forms of ∼4000 km$^2$ are relatively stable for ∼30 s during interval a compared to ∼20 s for the aurora in interval b. For large auroral forms the situation is turned around with the aurora in interval a having the most variable characteristics. Auroral forms of ∼10,000 km$^2$ are relatively stable for ∼50 s during interval a compared to ∼100 s for the aurora in interval b. Moving the threshold for what is considered relatively stable aurora to correlation values >0.7 would shift the times to lower values and slightly lower differences between the characteristics, but the overall findings are unaltered. These findings may be explained by our visual inspection used to select the time intervals of different auroral displays. It might be that this selection leads to a bias toward structures of a certain size, and we do not notice the larger structures that change on larger timescales. This calls for a more objective quantitative analysis than has been used in most past ASI studies, and it emphasizes the need for the robust technique used in this paper. Finally, it should not be overlooked that although the two intervals show differences they also show similarities.

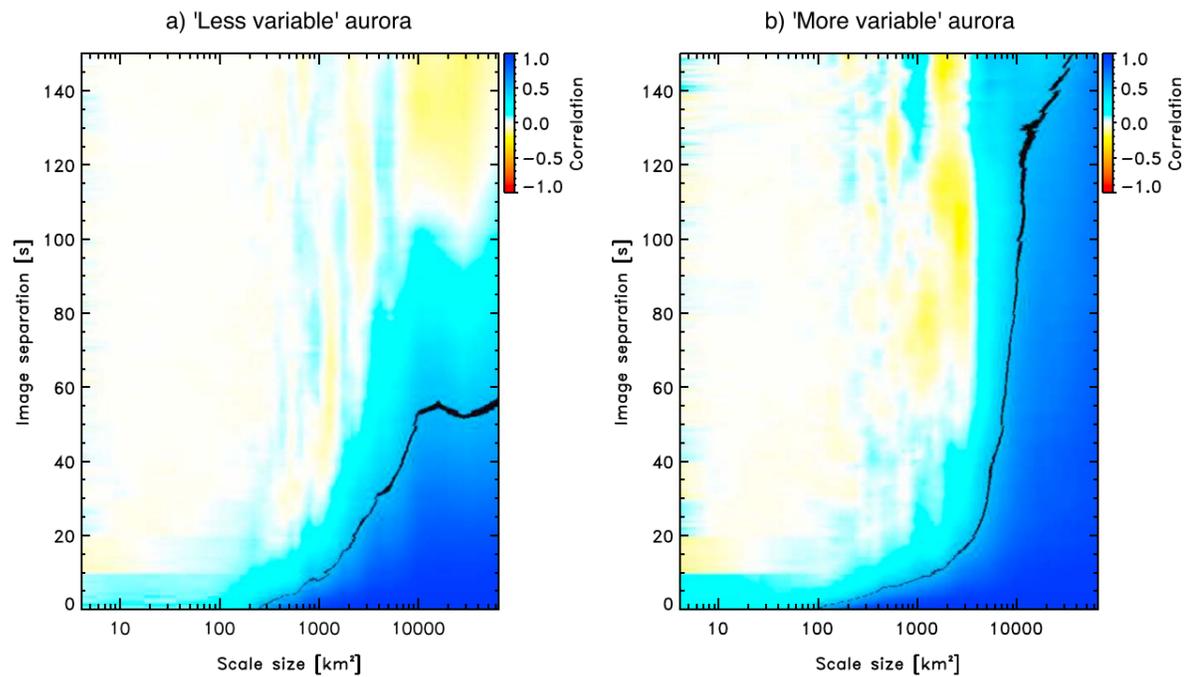

**Figure 7.** Correlation as a function of scale size and image separation (same format as Figure 5) for the intervals of (a) "less variable" and (b) "more variable" aurora. Correlation values of 0.5 are shown in black to elucidate the location of the correlated/uncorrelated boundary. We get an indication that the spatiotemporal characteristics of the aurora varies.

As we analyze emissions we are indirectly investigating the part of the precipitating electrons that produce the emissions. It thus seems reasonable that the characteristics of the emissions are also the characteristics of the precipitating electrons. For example, it seems reasonable that emissions produced by diffuse precipitation differs from those produced by Alfvénic or Inverted V precipitation. One of the differences between the intervals is that interval b includes diffuse aurora of a higher intensity over a larger part of the FOV and with more structure in the form of fluctuating/pulsating aurora compared to interval a. We note that the characteristics of the aurora in interval b is more variable for scale sizes typical for fluctuating/pulsating features ($\sim$1600–3600 km$^2$) and more stable for large scale sizes covering the relatively stable total area of the diffuse aurora, compared to the characteristics of interval a. We have, however, no objective way to separate the different emissions and their characteristics.

It is in general difficult to distinguish between the Inverted V/monoenergetic and Alfvénic/broadband precipitation by observing the discrete aurora alone. For example, *Colpitts et al.* [2013] observed Alfvénic aurora intermittently and perhaps simultaneously with inverted V aurora during an event of continuous substorm activity. They described the aurora as extremely dynamic changing on timescales of seconds with bright spots and arcs that were not as elongated in the east-west direction as during typical inverted V events. Neither interval a nor b clearly holds such features. Further, *Mende et al.* [2003a] found that in the substorm aurora outside of the surge, the Alfvénic electrons were less clearly separated from the inverted Vs, and *Mende et al.* [2003b] observed intense Alfvénic precipitation in the substorm auroral surge that were likely not present prior to the onset. It is therefore unfortunate that we could not study the substorm auroral bulge in the last minutes of data. If *Mende et al.* [2003b] are right, we could expect to see a difference in the lifetimes of different scale sizes of the distorted arcs before the substorm (monoenergetic) and the substorm auroral bulge (monoenergetic and Alfvénic aurora), which would enable a further discussion on how the characteristics vary in response to the types of electron precipitation that produce the discrete aurora.

We argue that our analysis is analogous to a study of for example field-aligned currents without a separation into underlying magnetospheric processes. As such our analysis could be applied to specific events such as north-south structures, poleward boundary intensifications and discrete arcs.

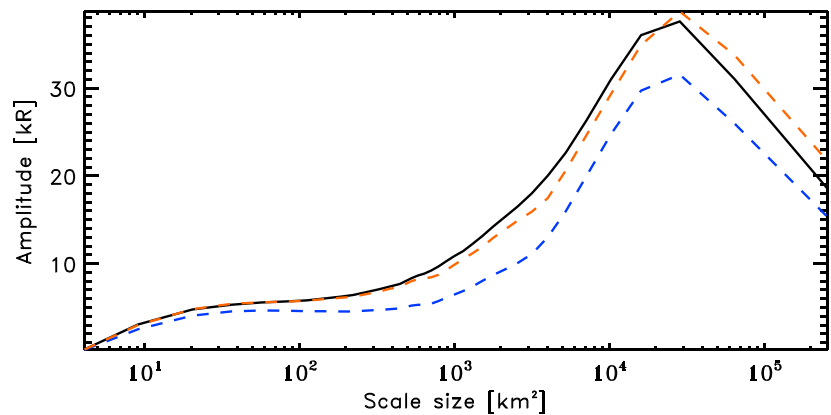

**Figure 8.** The scale size-dependent amplitude for all available images (black) and for the intervals of "less variable" (blue) and "more variable" (orange) aurora.

### 6.4. Significance of Different Scale Sizes

To understand the significance of the different auroral scale sizes, we determine the scale size-dependent amplitude $A = A(S)$. It is found from the complex spectrum as the square root of the band-pass power and thus tell us "how much" the band passed frequencies (and thus scale size) are contributing to the emission intensity in the original image. We find the amplitude for each band pass and then calculate the median scale size dependent amplitude over all available images. Figure 8 shows that the scale size of about 30,000 km$^2$ has the largest amplitude, while the amplitude falls off toward larger and smaller scale sizes. This corresponds to 173 km wide features when $s(f_x) = s(f_y)$. For scale sizes less than about 800 km$^2$ the amplitudes are less than 10 kR. The scale size-dependent amplitude is also found for the intervals of "less variable" (blue) and "more variable" (orange) aurora. The "more variable" aurora all scale sizes have a higher amplitude than the scale sizes of the "less variable" aurora. We can conclude that the smaller scale sizes both have shorter lifetimes and smaller amplitudes.

### 6.5. Comparison to the Spatiotemporal Characteristics of Field-Aligned Currents

Comparing our results to past findings is complicated by the observational challenges and resulting lack of statistical studies. However, [*Gjerloev et al.*, 2011], reported a study in which they determined the spatiotemporal characteristics of the magnetic field perturbations (dB) at low Earth orbit altitudes. The comparison is, however, not straightforward since the precipitating electrons producing the emissions are not necessarily the same that carry the current. This is obvious in the downward current region. Several past studies have tested if one can obtain field-aligned current (FAC) density by integrating the measured number flux of particles which is the definition of current density. Such a particle-based approach has been tested examining upward FACs collocated with accelerated auroral precipitation [e.g., *Hoffman et al.*, 1985; *Olsson et al.*, 1998; *Morooka et al.*, 2004]. It was found that FACs tend to be carried by high-energy electrons and an agreement with emissions would therefore be expected. However, the results showed significant disagreement between the current densities estimated from the auroral particle precipitation and from magnetic field measurements. This suggests the existence of missing current carriers, a problem that gets worse when the distance from the accelerated precipitation increases and obviously for downward currents.

Figure 9a shows the correlation as a function of dB (which we here interpret as indicative of FACs) scale size and time between the measurements from >4700 satellite crossings of the nightside auroral oval [*Gjerloev et al.*, 2011]. The time intervals between the measurements are determined by the variable interspacecraft separation of the three ST 5 satellites which were in a pearls-on-a-string formation (coorbital). Correlation coefficients of 0.5 are shown as black dots and used to fit the linear lines which elucidate the location of the correlated/uncorrelated boundary. The correlation value of 0.5 was an arbitrarily chosen value and a change to 0.6 would simply result in a shift in the positive *x* axis direction without any noteworthy change in the slope [*Gjerloev et al.*, 2011]. The correlations for satellite separations of about 6 s and scale sizes less than about 20 km are due to extrapolation since the dB was determined from spin-averaged measurements. The red line indicates the FAC filament scale size (1-D) of about 250 km, which square is the maximum auroral scale size (2-D).

Figure 9b is the result from Figure 5 for image separations of less than 200 s and average correlation values, as used for the spatiotemporal characteristics of the FACs. For easy comparison, the linear line of the ST 5 study is superposed as a dashed line and the correlation coefficients of about 0.5 are colored black. When comparing these lines, we find that the variability of the aurora and FACs follow each other closely until about 2000 km$^2$ (∼45 km). For larger scale sizes the aurora is more stable than the FACs. However, the displayed nightside characteristics of the FACs include both quiet and disturbed events, while the characteristics of the auroral emissions are from one moderately disturbed event (SML ≈ −300 nT). Therefore, a second relationship of the FAC characteristics during disturbed conditions (AL < −100 nT) is superposed as a dotted line (valid up to about $\Delta T = 70$ s). When comparing the lines during disturbed conditions (solid and dotted), we find that they follow each other remarkably well. It appears that the disturbed conditions follow our results better than the one for all conditions.

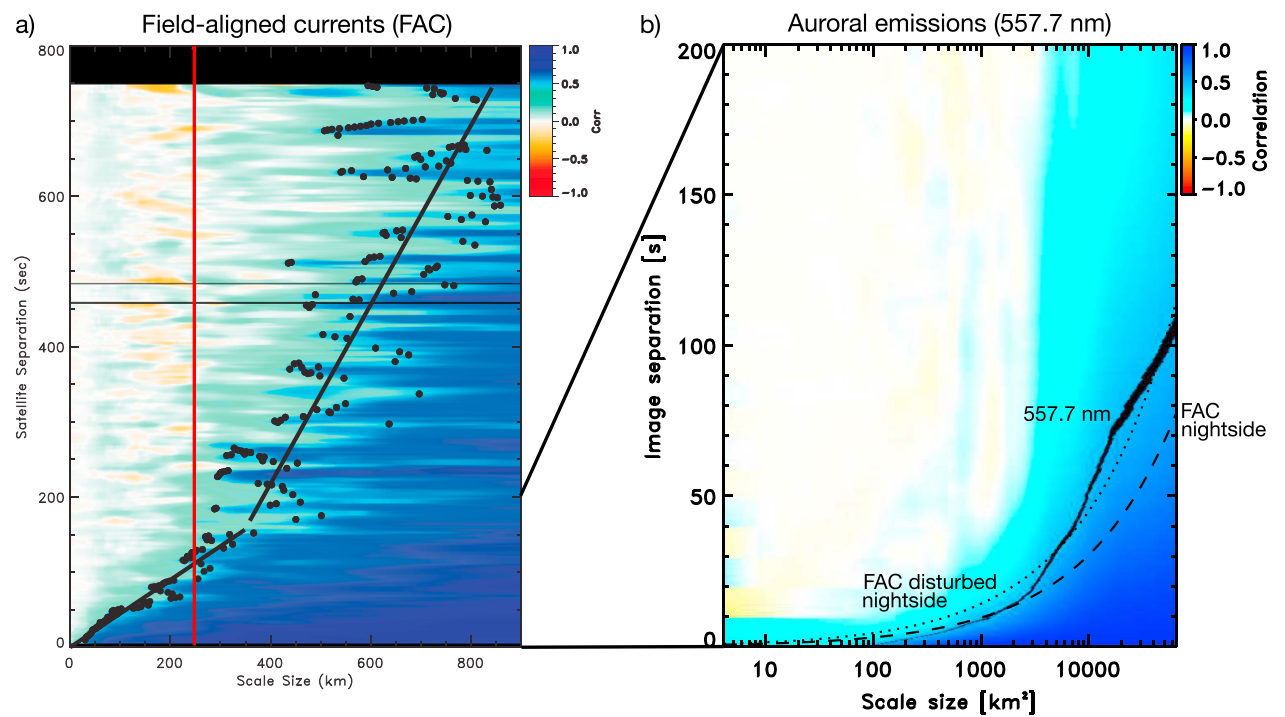

**Figure 9.** (a) Correlation as a function of FAC scale size and satellite separation for nightside events [*Gjerloev et al.*, 2011]. Correlation values of 0.5 are superposed (black dots) and used to fit the linear lines indicating the location of the correlated/uncorrelated boundary. The red line indicates the FAC filament scale size (1-D) of about 250 km, which square is the maximum auroral scale size (2-D). (b) The same as Figure 5 for image separations of less than 200 s showing the average correlation coefficients. The correlation coefficients of about 0.5 are colored black and the linear line of the ST 5 study is superposed as a dashed line for comparison. A second relationship of the FACs' characteristics during disturbed conditions (AL < −100 nT) is superposed as a dotted line. ăNote the remarkable agreement between the characteristics of the FACs and the auroral emission.

Despite the differences in data set, electromagnetic parameter, and technique, the resulting spatiotemporal characteristics of nightside auroral emissions and the nightside FACs are in remarkable agreement. At first glance, one may argue that the current carriers are also responsible for the emissions but as mentioned above this would be an oversimplification. Rather, we interpret this as indicative that the spatiotemporal characteristics of the emissions are similar to those of dB (or FACs). This seems logical since both are part of the M-I system and in fact it seems difficult to argue that they should differ. We do, however, still emphasize the striking similarity between these two studies despite the differences in technique and electrodynamic parameter.

## 7. Summary and Conclusion

We have determined objective and quantitative spatiotemporal characteristics of premidnight mesoscale auroral emission (557.7 nm) during a period of fairly constant moderate geomagnetic disturbances. The single event studied in this paper provides an opportunity to prove the technique and provides a glimpse of the M-I system characteristics, but it does not allow general conclusions. Below, we summarize the findings.

1. We find a scale size-dependent variability where the largest scale sizes are stable on timescales of minutes while the small scale sizes are more variable. For example, an auroral form of 1000 km$^2$ can be considered as static on timescales of about 10 s or less, while a 10,000 km$^2$ auroral form can be considered as static on timescales less than a minute.
2. We question the assumption that the spatiotemporal characteristics are fairly constant by separating the movie into two smaller intervals of different types of auroral displays. We find that the spatiotemporal characteristics varies during the event. The interval of less variable aurora has more stable auroral forms of

3. ~2000–5000 km$^2$ and less stable large auroral forms (>9000 km$^2$) compared to the interval of more variable aurora. However, the trend of increasing variability toward the smaller scale sizes is similar.
4. We have effectively found the scale size-dependent total derivative. However, an estimate shows that the source of the variability is mostly due to change in the partial derivative with respect to time and that we largely can ignore the convective term of the auroral arcs.
5. The scale size of about 30,000 km$^2$ is the most significant, while the amplitude falls off toward larger and smaller scale sizes. For scale sizes less than about 800 km$^2$ the amplitudes are less than 10 kR.
6. The average spatiotemporal characteristics of the auroral emissions are in remarkable agreement with the spatiotemporal characteristics of the nightside FACs during moderately disturbed times. Thus, two different electrodynamical parameters of the M-I coupling show similar behavior. This result is interpreted as an indication of a system that uses repeatable solutions to transfer energy and momentum from the magnetosphere to the ionosphere.

**Acknowledgments**
This study was supported by the Research Council of Norway under contract 223252. The authors acknowledge the use of SuperMAG indices and all-sky imager data from the Multi-spectral Observatory of Sensitive EMCCDs (MOOSE). The SuperMAG indices were obtained freely from supermag.uib.no. We greatly acknowledge James Weygand for the ACE solar wind data. MOOSE all-sky imager data were obtained from R.G. Michell and M. Samara. The data analyzed in this study are available upon request from the authors.